\documentclass[11pt,
 reprint,
superscriptaddress,
nofootinbib,
 amsmath,amssymb,
 aps,
]{revtex4-2}

\usepackage{booktabs}  
\usepackage{xcolor}    
\usepackage{array} 
\usepackage{graphicx}
\usepackage{dcolumn}
\usepackage{bm}
\usepackage{hyperref}
\usepackage{soul}
\usepackage{cancel}

\hypersetup{colorlinks,linkcolor=blue
}
\usepackage{amsmath}
\usepackage{amssymb}
\usepackage{amsthm}

\newcommand\be{\begin{equation}}
\newcommand\ee{\end{equation}}

\newcommand{\dd}{{\rm d}}

\newcommand\R{\mathcal{R}}

\begin{document}


\title{Spatial anisotropies from long wavelength scalar and tensor modes}

\author{Jorge Nore\~na}
\email{jorge.norena@pucv.cl}
\affiliation{Instituto de F\'isica, Pontificia Universidad Cat\'olica de Valpara\'iso, \\Casilla 4950, Valpara\'iso, Chile}
\author{Thiago S. Pereira}
 \email{tspereira@uel.br}
\affiliation{Departamento de F\'isica, Universidade Estadual de Londrina, Rod. Celso Garcia Cid, Km 380, 86057-970, Londrina, Paran\'a, Brazil}
\affiliation{Instituto de F\'isica, Universidade Federal do Rio de Janeiro, 21941-972, Rio de Janeiro, RJ, Brazil}
\author{Sean Reynolds}
\email{sean.reynolds.r@mail.pucv.cl}
\affiliation{Instituto de F\'isica, Pontificia Universidad Cat\'olica de Valpara\'iso, \\Casilla 4950, Valpara\'iso, Chile}


\begin{abstract}
We investigate the dominant physical effects of superhorizon fluctuations in a flat FLRW universe, focusing on whether the combined evolution of scalar and tensor adiabatic modes in the near-horizon regime could lead to geometries beyond those predicted by the conventional separate-universe approach. Assuming a matter-dominated universe and working to first order in perturbations but second order in a gradient expansion, we identify modes that are either pure gauge or unsourced, making them observationally irrelevant. This allows us to derive an effective metric that preserves the spatial symmetries of three well-known Bianchi cosmologies, namely, types I, V, and IX. In this framework, scalar perturbations induce spatial curvature, while the shear arises from long-wavelength tensor perturbations.
\end{abstract}

\maketitle


\section{Introduction}

A key ingredient of modern cosmology is the cosmological principle, according to which our universe is well approximated by a homogeneous and isotropic background spacetime, whose geometry is fully described by the Friedmann-Lemaître-Robertson-Walker (FLRW) metric. The inclusion of small perturbations governed by Einstein equations, plus the ad-hoc mechanism of an early single-field inflationary phase, lead to a mathematical model where the universe is described with just six free parameters (see~\cite{Anselmi:2022uvj} for a different point of view). The success of this description has been largely confirmed by CMB and large-scale structure data~\cite{Planck:2018vyg,eBOSS:2020yzd}. At the same time, there are also claims of large-angle CMB anomalies and tensions between low and high redshift data~\cite{Schwarz:2015cma,Planck:2015igc,Abdalla:2022yfr,Aluri:2022hzs}, suggesting either some subtle systematic effect in data analysis, or that we may have reached the limits of our theoretical description. This scenario has boosted intense theoretical activity in the past years, where basic ingredients are modified or new degrees of freedom are introduced to account for the anomalies and tensions~\cite{Bull:2015stt}.

However, a deeper scrutiny of the standard cosmological framework may reveal unexpected features even without the introduction of additional hypotheses. For example, backreaction effects are known to introduce spatial anisotropies that can affect the isotropy of the Hubble flow~\cite{Marozzi:2012ib}. More recently, it has been shown that anisotropic dynamics will emerge from the homogeneous (i.e., long-wavelength) limit of standard cosmological perturbations, and thus all cosmological observables are expected to show some deviation from isotropy at large scales~\cite{Pereira:2019mpp}.

The possible interplay between cosmological perturbations and background features is not new. In fact, it is well known that a horizon-sized density fluctuation on top of an otherwise flat background mimics the geometrical effects of spatial curvature~\cite{Brandenberger:2004ix,Geshnizjani:2005ce,Hirata:2005ei,Baldauf:2011bh,Kleban:2012ph}, and can even hinder our ability to constrain $\Omega_k$~\cite{Knox:2005hx,Waterhouse:2008vb,Leonard:2016evk,Tian:2020qnm,Blachier:2023ooc}. Likewise, a gravitational wave of infinite wavelength propagating on a flat background can mimic an anisotropic expansion~\cite{Talebian-Ashkezari:2016llx,Pereira:2019mpp} --- a property that has been leveraged to calculate weak-lensing observables in Bianchi spacetimes~\cite{Adamek:2015mna,Marcori:2018cwn}. If instead a gravitational wave of maximal wavelength is placed in a background with positive curvature, the resulting metric is that of a Bianchi-IX universe~\cite{King:1991jd}. This result is perhaps an extreme example of this perturbation/background duality, since both the gravitational wave and the curvature are treated non-perturbatively. 

Motivated by these works, we ask whether scalar and tensor perturbations of long wavelengths (but small amplitudes) could conspire to produce homogeneous spacetimes with known symmetries. In other words, will the scalar-induced spatial curvature combine with tensor-induced shear to preserve the symmetries of some (spatially curved) Bianchi cosmology? At first sight, the answer seems to be negative since, working at second order in a gradient expansion, the Hessian matrix of scalar and tensor perturbations contains off-diagonal terms with no obvious symmetries. However, we show that these terms will not play a role if we assume adiabatic matter perturbations in comoving gauge. As a result, we derive an effective metric that is fully described by the (long-wavelength) curvature and tensor perturbations. As it turns out, this metric has the same symmetries of Bianchi metrics of types I, V, and IX, according to whether the curvature is negative, zero, or positive.

Our results differ from previous and related works~\cite{King:1991jd,Pontzen:2010eg,Pereira:2019mpp} in two important aspects. First, rather than showing that Bianchi universes with small anisotropies are equivalent to perturbed FLRW metrics, we ask about the leading physical effect of superhorizon modes in a flat FLRW universe with adiabatic perturbations, and we show that it is locally described by some Bianchi metrics. Indeed, we find that not all Bianchi solutions with a FLRW limit can arise in such a setting.
This is motivated by current CMB observations which show that perturbations are compatible with adiabatic evolution and $\Omega_k$ is consistent with zero and very close to the curvature floor of $10^{-4}$~\cite{Leonard:2016evk}. Second, our derivation is essentially geometrical. Although we use Einstein equations as a no-go theorem for some Bianchi models, our main results derive essentially from an effective metric and its symmetries.

We work by expanding the metric perturbations in gradients, since we are interested in long-wavelength modes. The leading and subleading order in this expansion has no physical effect on subhorizon scales, while the leading physical effect of a scalar perturbation is to induce a small curvature. This is reviewed in section \ref{sec:gradient_expansion}. We then ask what is the leading physical effect of tensor perturbations in the presence of spatial curvature, which turns out to be described by a subset of Bianchi solutions, so these are presented in section \ref{sec:bianchi}. We show this equivalence in section \ref{sec:main}. Finally, we present our conclusions and future directions in section \ref{sec:conclusions}. Some ancillary calculations are relegated to the Appendixes.

\section{Gradient expansion}
\label{sec:gradient_expansion}

We shall consider metric perturbations in comoving gauge sourced by a single-field inflationary model. To linear order in perturbations (which should be a good approximation for superhorizon modes), the metric reads~\cite{Maldacena:2002vr}
\begin{multline}\label{eq:starting-metric}
ds^2 = -(1 + 2\delta N)dt^2 + 2 a^2 N_i dx^i dt \\ + a^2(t)\left[(1 + 2\R)\delta_{ij} + \gamma_{ij}\right]dx^i dx^j\,,
\end{multline}
where $\R$ and $\gamma_{ij}$ are the curvature and tensor perturbations, respectively, and $\gamma^i_i = 0 = \partial_i\gamma_{ij}$. Latin indices are raised and lowered with $\delta_{ij}$. The shift vector $N_i$ is decomposed into a longitudinal and transverse part as $N_i = \partial_i \psi  + N_i^{\perp}$, with $\partial^i N_i^\perp = 0$. The lapse perturbation $\delta N$ and the shift $N_i$ can be written in terms of $\R$ using the Hamiltonian and momentum constraint equations as (see Appendix~\ref{sec:constraints})
\begin{equation}
  \delta N = \frac{\dot{\R}}{H}\,,\quad N_i \sim \mathcal{O}(q/H)\,.
  \label{eq:lapse_shift_solution}
\end{equation}
where $q$ is a Fourier mode.

We focus on superhorizon perturbations (which can be approximated as a constant in space at subhorizon scales). For that we expand $\R$ and $\gamma_{ij}$ around $\mathbf{x}=\mathbf{0}$ to second order in gradients, which gives
\begin{align}
\begin{split}\label{eq:grad-expansion}
    \R(t, \mathbf{x}) & \approx \R(t) + x^k\partial_k\R + \frac{1}{2}x^kx^l\partial_k\partial_l\R\,,\\
    \gamma_{ij}(t, \mathbf{x}) & \approx 
    \gamma_{ij}(t) + x^k\partial_k\gamma_{ij} + \frac{1}{2} x^kx^l\partial_k\partial_l\gamma_{ij}\,,
\end{split}
\end{align}
where $x^k\partial_k(\cdots\,\!)$ is a shortcut for $x^k\left[\partial_k (\cdots\,\!)|_{\mathbf{x}=\mathbf{0}}\right]$, and so on.

For an adiabatic evolution, the curvature fluctuation $\R$ is constant outside the horizon~\cite{Bardeen:1980kt, Lyth:1984gv}, its time derivative evolving as $\dot{\R} \sim \mathcal{O}(q^2/H^2)$~\cite{Bardeen:1980kt}. Similarly, if the background evolution is an isotropic attractor (i.e. if anisotropy decays with time), tensor modes are constant outside the horizon~\cite{Bordin:2016ruc}, and we also have $\dot{\gamma}_{ij} \sim \mathcal{O}(q^2/H^2)$. Thus, at second order in the gradient expansion, only the leading terms in~\eqref{eq:grad-expansion} can evolve with time.

In what follows, we quickly will review the fact that, when working to first order in the gradient expansion, the leading and sub-leading terms are just coordinate artifacts. In fact, physical gravitational effects should be proportional to $\mathbf{x}^2$, as can be explicitly shown using Conformal Fermi Coordinates \cite{Dai:2015rda}. One can find coordinates which satisfy the comoving gauge conditions where those terms are absent \cite{Creminelli:2012ed, Hinterbichler:2012nm}. At sub-sub-leading order, we re-derive the known fact that the \emph{isotropic} curvature fluctuation (defined by the Laplacian $\nabla^2\R$) is locally equivalent to a curved background (characterized by the spatial curvature $k$). Therefore, in the presence of such a perturbation, an observer will measure a slightly curved universe. Based on these results, we then ask if something analogous happens to other types of sub-sub-leading terms. This will lead to the metric~\eqref{eq:metric}, which is the basis of our main findings.

\subsection{Constant metric fluctuations (leading order)}

At leading order in the gradient expansion, $\R$ and $\gamma_{ij}$ are pure constants, and we can ignore the lapse and the shift. At this order the metric reads
\be
ds^2=-dt^2 + a^2(t)\left[(1+2\R)\delta_{ij}+\gamma_{ij}\right]dx^idx^j\,.
\ee
It is straightforward to show that this is just the flat FLRW metric after performing a linear coordinate transformation,
\begin{equation}
  x^i \mapsto x^i(1 - \R) - \frac{1}{2}\gamma_{ij}x^j\,.
  \label{eq:constant_transf}
\end{equation}
Thus, these perturbations have no effect on subhorizon physics~\cite{Maldacena:2002vr, Creminelli:2004yq}.

\subsection{Constant gradient (sub-leading order)}

Constant gradients of metric fluctuations can also be eliminated by a change of coordinates \cite{Creminelli:2012ed, Hinterbichler:2012nm}. Let us first consider the curvature perturbation which, at this order, 
can still be taken as constant in time, though $N_i$ can no longer be ignored. The metric then becomes
\be
ds^2= -dt^2 + 2a^2(t)N_idx^idt + a^2(t)(1+2 x^i \partial_i \R)d\mathbf{x}^2\,.
\ee
The transformation that eliminates the gradient is simply
\be
x^i \mapsto x^i - 2(b^jx_j)x^i + b^ix^jx_j - \delta x^i(t)\,,
\ee
with $b_i = \partial_i\R/2$, and the time-dependent shift $\delta x^i$ is taken to be such that $N^i = - d\delta x^i(t)/dt$ \cite{Creminelli:2012ed}.

Now, consider a gradient of a tensor perturbation. That is
\be
ds^2=-dt^2 + a^2(t)(\delta_{ij} + x^k\partial_k \gamma_{ij})dx^idx^j .
\ee
This can also be eliminated by a coordinate transformation \cite{Hinterbichler:2012nm}, which is \footnote{Note that this is also the transformation to the Riemann coordinates on the constant $t$ hypersurface at this order.}
\be
x^i \mapsto x^i - \frac{1}{4}(\partial_j \gamma_{ik} + \partial_k \gamma_{ij} - \partial_i \gamma_{jk})x^j x^k\,.
\ee 

\subsection{A Laplacian of the curvature fluctuation is locally equivalent to a curved universe}
\label{sec:laplacian_curvature}

It is well known that the Laplacian of the curvature perturbation is locally equivalent to the spatial curvature 
(for a modern reference, see \cite{Baldauf:2011bh,Blachier:2023ooc}.) This result is not trivial, since the Laplacian also introduces a $\mathbf{x}^2$ term that makes the metric inhomogeneous. To see how this is possible in principle, we recall that the curved FLRW metric in Cartesian coordinates is~\cite{weinberg2008cosmology}
\begin{equation}\label{eq:FLRW-metric}
ds^2_\text{FLRW} = -dt^2 + a^2(t)\left[ d\mathbf{x}^2 + \frac{k (\mathbf{x}\cdot d\mathbf{x})^2}{1-k\mathbf{x}^2} \right]\,.
\end{equation}
At linear order in the curvature $k$, and after a change of coordinates $x^i \mapsto x^i-\frac{1}{4}kx^i\mathbf{x}^2$, it reads
\be\label{eq:curved_FLRW_small_k}
ds^2_\text{FLRW} \approx -dt^2 + a^2(t) \left( 1 - \frac{1}{2}k \mathbf{x}^2 \right)d\mathbf{x}^2\,.
\ee
Note the similarity between the quadratic term and the isotropic component, $\mathbf{x}^2\nabla^2\R/3$, which originates from the trace of the Hessian matrix, $\partial_k\partial_l\R$, in the expansion \eqref{eq:grad-expansion}. However, there are two remaining subtleties if we want to derive \eqref{eq:curved_FLRW_small_k} from \eqref{eq:starting-metric}. First, recall that, at second order in gradients, we can no longer ignore the time dependence of the leading term in \eqref{eq:grad-expansion} (and thus the lapse \eqref{eq:lapse_shift_solution}). Second, the gradient expansion also introduces anisotropic derivatives of $\R$ which do not appear in the metric above. We shall come back to this issue later, and for the moment we just focus on the diagonal part. Therefore, the perturbed metric \eqref{eq:starting-metric} takes the form
\begin{multline}
ds^2 = -\left(1+2\frac{\dot{\R}}{H}\right)dt^2 \\ + a^2(t)\left(1+2\R(t) + \frac{1}{3}\mathbf{x}^2\nabla^2\R \right)d\mathbf{x}^2\,,
\label{eq:perturbed_metric_laplacian}
\end{multline}
where we have ignored the time-dependence of $\partial_i \R$ and $\nabla^2 \R$ (since they would be of higher order in $q/H$), and we eliminated the gradient through the coordinate transformation of the previous section.

One difference between the metrics \eqref{eq:curved_FLRW_small_k} and \eqref{eq:perturbed_metric_laplacian} is that the latter is not in a synchronous coordinate system. This is easily fixed with a time transformation $t \mapsto t + \xi^0$ with
$$
\dot{\xi}^0=-\frac{\dot{\R}}{H}\,.
$$
The metric then becomes
\be
ds^2 = -dt^2 + b^2(t) \left( 1 - \frac{1}{2}k \mathbf{x}^2 \right)d\mathbf{x}^2\,,
\label{eq:curved_flrw_with_b}
\ee
where 
\begin{equation}\label{eq:b-k-def}
b(t)\equiv a(t+\xi^0)(1+\R(t))\,,\quad  k \equiv -\frac{2}{3}\nabla^2\R\,.
\end{equation}
This has the same form as the curved FLRW solution \eqref{eq:curved_FLRW_small_k}. For consistency, we finally check in Appendix \ref{sec:einstein_laplacian} that the function $b(t)$ satisfies the Friedmann equations of the curved universe. 

In the derivation above we have ignored tensor perturbations. As we show next, their inclusion in~\eqref{eq:curved_flrw_with_b} will produce a metric corresponding to a subclass of Bianchi spacetimes, which are spatially homogeneous cosmological solutions allowing the universe to expand anisotropically. Since this correspondence is not trivial, we quickly review the basic ingredients in the construction of anisotropic Bianchi solutions, which will then be used to complete our proof. For further details, see Refs.~\cite{gron2007einstein,ellis2012relativistic,plebanski2024introduction}

\section{Bianchi models in a nutshell}
\label{sec:bianchi}
We start by recalling some basic facts on continuous symmetries of spaces with metric $g$. Such symmetries, or isometries, are transformations of points that preserve the metric; in other words, these are transformations generated by Killing vector fields $\xi_i$:
\begin{equation}\label{eq:Lie-g}
    {\cal L}_{\xi_i}{g} = 0\,,
\end{equation}
where ${\cal L}$ represents a Lie derivative. In a space of dimension $d$, the maximum number of independent Killing vectors is $d(d+1)/2$, of which exactly $d$ generate translations, while the remaining $d(d-1)/2$ generate rotations~\cite{Weinberg:1972kfs}. Moreover, since in $d$ dimensions there can only be $d$ linearly independent vectors, one can show that the $d(d-1)/2$ rotation vectors are position-dependent combinations of the translation vectors, and thus are not everywhere linearly-independent. A simple example is given by a rotation in flat space, say, $x\partial_y - y\partial_x$.  Thus, we define Bianchi spaces as those having 3 Killing vectors which are everywhere linearly-independent, also known as simply transitive spaces. Since these isometries form a Lie group, their generating vectors form a Lie algebra, and we can classify all Bianchi spaces by finding all possible Lie algebras of Killing vectors $\xi_i$: 
\begin{equation}
[\xi_i,\xi_j] = C^k_{ij}\xi_k\,,\qquad (i=1,2,3)\,.
\end{equation}
In other words, we can complete the task by finding all \emph{constants of structure} $C^{k}_{ij}$. These can be written as 
\begin{equation}
C^k_{ij} = \epsilon_{ijl}n^{lk} - 2a_{[i}\delta^k_{j]}\,,\qquad n^{ij}a_j=0\,,
\end{equation}
where $n^{ij}$ and $a_i$ are a symmetric matrix and a vector, whose orthogonality follows from the Jacobi identity. With suitable linear combinations of the vectors $\xi_i$, one can write $n^{ij}=\text{diag}(n^1,n^2,n^3)$ and $a_i=(0,0,a)$ (see \cite{plebanski2024introduction} for details). Thus, the task is completed by tabulating all possible values of these four quantities. We show in Table~\ref{tab:bianchi-types} all Bianchi models with a FLRW limit, including those considered in this work.
\begin{table}[h]
    \centering
    \begin{tabular}{>{\bfseries}c c c c c c}
        \toprule
        Type & $a$ & $n^1$ & $n^2$ & $n^3$ & $k$\\
        \midrule
        I       & 0 & 0 & 0 & 0 & $=0$ \\
        V       & $1$& 0 & 0 & 0 & $<0$ \\
        VII$_0$ & 0 & $1$& $1$& 0 & $=0$ \\
        VII$_h$ & $\sqrt{h}$& $1$& $1$& 0 & $<0$ \\
        IX      & 0 & $1$& $1$& $1$& $>0$ \\
        \bottomrule
    \end{tabular}
    \caption{Table of cosmological Bianchi types with isotropic limit. $h$ is a constant (real) parameter, and $k$ is the curvature of the underlying FLRW space (see Eq.~\eqref{eq:FLRW-metric}).}
    \label{tab:bianchi-types}
\end{table}

The discussion above is restricted to spatial manifolds. Now we would like to use them to build cosmological solutions that (i) have the symmetries of Bianchi spaces, and (ii) preserve these symmetries as the universe evolves in time. The second condition is achieved by introducing a family of Bianchi spaces, each labeled by a continuous parameter $t$, and such that $e_0\cdot\xi_i=0$, where ${e}_0=-\partial_t$.

For the first condition, we introduce a basis of vectors $e_i$ such that $e_i\cdot e_j=g_{ij}(t)$ and $e_0\cdot e_i=0$. Condition~\eqref{eq:Lie-g} then translates to
\begin{equation}\label{eq:invariant-basis-def}
{\cal L}_{\xi_i}e_j = 0 = [\xi_i,e_j]\,,
\end{equation}
that is, the vectors $e_i$ should be invariant when spatially dragged along the isometries of the space. If the Killing vectors are explicitly known, one can directly solve Eq.~\eqref{eq:invariant-basis-def}, using $e_i(p)=\xi_i(p)$ for some point $p$, to find the basis $e_i$\footnote{Alternatively, one can show that 
$[e_i,e_j] = - C^k_{ij}e_k$. This then allows a direct computation of the dual basis from Cartan's structural equations: $\dd e^{k}=\frac{1}{2}C^{k}_{ij}e^i\wedge\,e^j$, where $\dd$ is the exterior derivative operator.}. From the solution, one easily constructs the dual basis $e^{i}$ from $e^{i}\cdot e_{j}=\delta^i_j$, which finally gives the metric as
\begin{equation}\label{eq:spacetime-metric}
ds^2 = -dt^2 + g_{ij}(t)e^ie^j\,.
\end{equation}
where $-dt = e^0$ is the dual vector to $e_0$. 

Given the metric \eqref{eq:spacetime-metric}, we can compute Einstein equations for a given source. In this work we are interested in a perfect and \emph{non-tilted} fluid, for which
\[
T^{\mu}_{\;\nu} = \text{diag}(-\bar{\rho},\bar{p},\bar{p},\bar{p})\,,
\]
where $\bar{\rho}$ and $\bar{p}$ are the fluid's energy density and pressure as measured in the frame defined in~\eqref{eq:spacetime-metric}, respectively, and which depend only on time. It is also convenient to write $g_{ij}(t)=a^2(t)\gamma_{ij}(t)$, where $a(t)$ is the average scale factor.  We then find, after linearizing over 
$\gamma_{ij}$~\cite{Pereira:2019mpp,Pontzen:2010eg}
\begin{align}
H^2 & = \frac{8\pi G}{3} \bar{\rho} - \frac{k}{a^2} + {\cal O}(\gamma^2)
\label{eq:G00_bianchi}\\
2\dot{H} + 3H^2 & = -8\pi G \bar{p} - \frac{k}{a^2} + {\cal O}(\gamma^2) 
\label{eq:G0i_bianchi}\\
\ddot{\gamma}_{ij} + 3H\dot{\gamma}_{ij} & = a^{-2}\left[2n^k_k n_{\langle ij\rangle} - 4n_{k\langle i}n^{k}_{j \rangle} + 4a^k \epsilon_{kl\langle j}n^l_{i\rangle}\right]
\label{eq:Gij_bianchi}
\end{align}
where indices enclosed by $\langle\dots\rangle$ means that a tensor is totally symmetric and trace-free, and where $k$ is given by the three-dimensional Ricci
\begin{equation}
    k = - a_i a^i + \frac{1}{6}\left(-n^{ij}n_{ij} + \frac{1}{2}(n^i_i)^2\right)\,.
    \label{eq:curvature_bianchi}
\end{equation}

\section{Second derivatives of metric fluctuations}
\label{sec:main}

Moving forward, we now prove that a perturbed FLRW solution is locally equivalent to a Bianchi universe. That is, keeping terms up to $\mathcal{O}(\nabla^2)$ in the gradient expansion of tensor and scalar perturbations, we show that the metric \eqref{eq:starting-metric} is indistinguishable from a subset of Bianchi metrics with small curvature and anisotropies, namely, models I, V and IX (see Table~\ref{tab:bianchi-types}).

\subsection{Metric}

After performing the coordinate transformations that eliminate the leading and sub-leading orders in the gradient expansion (that is, a constant $\R$ and $\gamma_{ij}$, plus their gradients), and now including both off-diagonal gradients of $\R$ as well as the tensor modes, we are left with the metric
\begin{multline}
ds^2 = -\left(1 + 2\frac{\dot\R}{H}\right)dt^2 + a^2(t)\Big[(1 + 2\R(t))\delta_{ij} \\ + \gamma_{ij}(t) + A_{ijkl}x^kx^l\Big]dx^idx^j\,,
\end{multline}
where $A_{ijkl} = \partial_k\partial_l\R \delta_{ij} + \frac{1}{2}\partial_k \partial_l \gamma_{ij}$ and, once again, we stress that we cannot ignore the time-dependence in $\R$ and $\gamma_{ij}$.

At second order in powers of $\mathbf{x}$, there are 10 adiabatic modes (i.e., coordinate redundancies) \cite{Hinterbichler:2013dpa}, see \cite{Mirbabayi:2014zpa} for the counting. Therefore, it is not a priori clear which coordinates are physically meaningful.\footnote{This is also true in conformal Fermi coordinates, which have a residual redundancy in the spatial part of the metric at this order in $\mathbf{x}$.} To circumvent this issue, we first isolate the $\mathcal{O}({\bf{x}}^2)$ terms in the spatial metric by performing the following transformation
\be
t \mapsto t + \xi^0\,,\quad x^i \mapsto x^i - \frac{1}{2}\gamma_{ij}x^j\,,
\ee 
with $\dot{\xi^0} = -\dot\R/H$. The metric then becomes
\begin{multline}
ds^2 = -dt^2 - b^2(t)\dot\gamma_{ij}(t)x^jdx^idt \\+ b^2(t)\Big(\delta_{ij} + A_{ijkl}x^kx^l\Big)dx^idx^j\,,
\end{multline}
where the scale factor $b(t)$ was defined in~\eqref{eq:b-k-def}. In order to isolate the curvature modes of the constant-$t$ hypersurface, we go to Riemann coordinates, which gives the following induced metric
\be
g_{ij}^{(3)}= \delta_{ij}+ \!\!\phantom{.}^{(3)}\!R_{ikjl}x^kx^l\,,
\ee
where $\!\!\phantom{.}^{(3)}\!R_{ikjl}$ is the three-dimensional Riemann tensor, and we have chosen units to reabsorb the scale factor. This tensor is completely fixed in terms of the three-dimensional Ricci tensor (see e.g. \cite{Weinberg:1972kfs})
\begin{multline}
\!\!\phantom{.}^{(3)}\!R_{ikjl} = \delta_{ij}\!\!\phantom{.}^{(3)}\!R_{kl}-\delta_{il}\!\!\phantom{.}^{(3)}\!R_{jk}-\delta_{jk}\!\!\phantom{.}^{(3)}\!R_{il}\\+\delta_{kl}\!\!\phantom{.}^{(3)}\!R_{ij}  - \frac{\!\!\phantom{.}^{(3)}\!R}{2}(\delta_{ij}\delta_{kl}-\delta_{il}\delta_{jk})\,,
\end{multline}
where we have used the fact that $\!\!\phantom{.}^{(3)}\!R_{ij}$ is linear in perturbations.
The induced metric then takes the form \footnote{Notice that, in this form, the traceless part is not transverse. If we denote $\tilde{g}_{ij}^{(3)}$ as the traceless piece of the metric, its divergence is
$$
\partial_i\tilde{g}_{ij}^{(3)}=\frac{5}{9}\!\phantom{.}^{(3)}\!R_{ij}x^i\neq 0 .
$$
}
\begin{multline}
g_{ij}^{(3)} =\delta_{ij} + \delta_{ij}\!\!\phantom{.}^{(3)}\!R_{kl} x^k x^l-\!\!\phantom{.}^{(3)}\!R_{jk} x^i x^k-\!\!\phantom{.}^{(3)}\!R_{ik} x^j x^k \\ +\!\!\phantom{.}^{(3)}\!R_{ij} \mathbf{x}^2 - \frac{\!\!\phantom{.}^{(3)}\!R}{2}(\delta_{ij}\mathbf{x}^2- x^i x^j)
\,.
\end{multline}
One can find a coordinate transformation that makes this metric a pure trace. In fact, take \footnote{The general transformation that gets rid of the longitudinal part of the traceless piece is
\begin{multline*}
x^i \mapsto x^i + \frac{1}{11}\left(\frac{1}{8} + 2\alpha\right)\!\phantom{.}^{(3)}\!Rx^i\mathbf{x}^2 \\ - \frac{1}{11}\left(\frac{5}{6} + 17\alpha\right)\!\phantom{.}^{(3)}\!R_{kl}x^kx^lx^i + \alpha \!\phantom{.}^{(3)}\!R^i_kx^k\mathbf{x}^2\,,
\end{multline*}
for any constant $\alpha$.
}
\be
x^i \mapsto x^i+\frac{1}{24}\!\phantom{.}^{(3)}\!Rx^i\mathbf{x}^2-\frac{1}{3}\!\phantom{.}^{(3)}\!R_{kl}x^kx^lx^i+\frac{1}{6}\!\phantom{.}^{(3)}\!R^i_kx^k\mathbf{x}^2\,.
\ee
Then the induced metric takes the form
\be
g_{ij}^{(3)}=\delta_{ij}\left(1 + \frac{1}{4}\!\phantom{.}^{(3)}\!R\mathbf{x}^2-\!\phantom{.}^{(3)}\!R_{kl}x^kx^l \right) + \mathcal{O}(\mathbf{x}^3)\,.
\ee
Therefore, we have reduced the metric to the simple form
\begin{multline}
ds^2 = -dt^2 - b^2(t)\dot\gamma_{ij}x^j dx^i dt \\+ b^2(t)\left(1 + A_{kl}x^kx^l\right)d\mathbf{x}^2\,.
\end{multline}
where $A_{kl} \equiv \frac{1}{4}\!\phantom{.}^{(3)}\!R\delta_{kl} - \!\phantom{.}^{(3)}\!R_{kl}$. To compare with Bianchi spacetimes, we take it to synchronous coordinates ($x^i \mapsto x^i - \gamma_{ij}x^j/2$)
\be
ds^2 = -dt^2 + b^2(t)\left[(1 + A_{kl}x^kx^l)\delta_{ij} + \gamma_{ij}(t)\right]dx^i dx^j\,.
\label{eq:eff-metric}
\ee
This has the shape of a perturbed FLRW universe in synchronous gauge with scale factor $b(t)$. However, in a matter-dominated universe, the Einstein equations source only specific forms of the function $A_{kl}$. Indeed, writing the 
space-space part of the Einstein tensor for this metric, and keeping only the contribution form the scalar perturbations, we find
\begin{equation}
    b^2\delta G^i_j = \left(-A^i_j + \frac{1}{3}A^k_k \delta^i_j\right) + \frac{2}{3}A^k_k \delta_{ij}\,,
\end{equation}
where we have separated the trace from the traceless part. In the absence of anisotropic stress, the traceless part of the space-space components of the energy-momentum tensor vanish, such that we are left with $A_{kl} = -\frac{k}{2}\delta_{kl}$. Here, we have used the fact that this trace is the locally measured curvature, see section \ref{sec:laplacian_curvature}.

Finally, the metric for a adiabatic linear perturbations in a matter-dominated universe, and at sub-sub-leading order in the gradient expansion, can be written as
\begin{equation}\label{eq:metric}
    ds^2 = -dt^2 + b^2(t)\left[\left(1 - \frac{k}{2}\mathbf{x}^2\right)\delta_{ij} + \gamma_{ij}(t)\right]dx^idx^j\,.
\end{equation}
Therefore, it is locally described by a homogeneous tensor mode superposed on a curved universe. From now on we drop the difference between the scale factors $a(t)$ and $b(t)$.

We will now show that this is indistinguishable from a Bianchi universe at this order in the expansions involved. We first show that the Einstein equations are compatible (see also \cite{Pereira:2019mpp}), and then we compare the metrics.

\subsection{Comparison of the Einstein tensor}
\label{sec:einstein_comparison}

The perturbed Einstein tensor for the metric \eqref{eq:metric} is
\begin{align}
  a^2\delta G^{i}_{\phantom{i}j} &= -k\delta^{i}_{j} + \ddot{\gamma}^{i}_{\;\;\; j} + 3H\dot{\gamma}^{i}_{\;\; j}\,,\label{eq:delta_Gij_iso}\\ 
  a^2\delta G^0_{\; i} &= 0\,,\label{eq:delta_Gi0_iso}\\ 
  a^2\delta G^0_{\; 0} &= -3k\,\label{eq:delta_G00_iso}.
\end{align}
We now compare this to the Einstein equations for the Bianchi solutions from section \ref{sec:bianchi}. Specifically, if we compare Eq.~\eqref{eq:delta_G00_iso} with Eqs.~\eqref{eq:G00_bianchi} and \eqref{eq:curvature_bianchi}, Eq.~\eqref{eq:delta_Gi0_iso} with \eqref{eq:G0i_bianchi}, and the traceless part of Eq.~\eqref{eq:delta_Gij_iso} with \eqref{eq:Gij_bianchi}, we find that they have the same form if
\begin{align}
  -a^ia_i + \frac{1}{6}\left(- n^{ij}n_{ij} + \frac{1}{2}\left(n_i^i\right)^2\right) & = k\,,\label{eq:conds_bianchi_00}\\
  3aa^j\dot\gamma_{ij} + \epsilon_{ijk}\gamma^{jl}n_l^k & = 0\,,\label{eq:conds_bianchi_0i}\\
  n^k_k n_{\langle ij \rangle} - 2 n_{k\langle i}n^k_{j\rangle} + 2 a^k \epsilon_{kl\langle j}n^l_{i\rangle} & = 0\,.\label{eq:conds_bianchi_ij}
\end{align}

Now, compare these with Table \ref{tab:bianchi-types}. From this comparison we conclude that 

\begin{itemize}
  \item When $k$ is zero, the metric is clearly that of a Bianchi I model, since it can be written as
  \begin{equation}
      ds^2 = -dt^2 + a^2(t)(\delta_{ij} + \gamma_{ij})dx^idx^j\,.
      \label{eq:my_bianchi_I}
  \end{equation}
We simply take \(a_i = 0 = n_{ij}\).
  
  \item When $k$ is negative, we can take \(a_i\) to be orthogonal to \(\gamma_{ij}\) and \(n_{ij} = 0\). That is, we take \(a_i\) in the direction to which \(\gamma_{ij}\) is transverse, making it compatible with equation \eqref{eq:conds_bianchi_0i}. Since \(a^i a_i\) is positive, this is compatible with equation \eqref{eq:conds_bianchi_00}. Therefore, Einstein equations are compatible with those of Bianchi V.
  
  \item When $k$ is positive, we can take \(n_{ij} \propto \delta_{ij}\) and \(a_i = 0\). In this case, Eqs. \eqref{eq:conds_bianchi_0i} and \eqref{eq:conds_bianchi_ij} are automatically satisfied. Moreover, Eq. \eqref{eq:conds_bianchi_00} is also satisfied since the left-hand side is negative. Therefore, the Einstein equations are the same as those of Bianchi IX.
  
  \item From Table~\ref{tab:bianchi-types}, we see that model VII$_0$ always satisfies conditions \eqref{eq:conds_bianchi_00} and \eqref{eq:conds_bianchi_ij}. Indeed, $n^{ij}n_{ij} + \frac{1}{2}(n^i_i)^2 = 0$ and $n^k_k n_{ij} - 2 n_{ki} n^k_j = 0$ identically for this model. We then only need to satisfy condition \eqref{eq:conds_bianchi_0i}, which is obeyed if we take $\gamma_{ij}$ to be transverse to the $3$ direction. However, from the explicit form of the metric in the flat case, we know that it corresponds to Bianchi I rather than VII$_0$. This shows that under our approximations (linear perturbations and gradient expansion) and in this setting (adiabatic evolution, and a perfect non-tilted fluid) the compatibility of Einstein's equations does not allow us to conclude that the metric is the same as that of a specific Bianchi solution.
  
  \item For model VII$_h$, one also needs to check the last term in condition \eqref{eq:conds_bianchi_ij}. The comparison fails in this case since it gives something proportional to $-2k\epsilon_{3ij}$, which is non-zero. Therefore, Bianchi VII$_h$ cannot arise as adiabatic perturbations of a FLRW universe. This makes sense since model VII$_h$ involves an additional spiraling scale which is not present in our perturbative expansion (see~\cite{Pereira:2019mpp}).
\end{itemize}

\subsection{Comparison of the metric}
\label{sec:metric_comparison}

We are now in position to compare the perturbed metric \eqref{eq:metric} with that of a Bianchi universes. Bianchi metrics are defined by its symmetries, which are associated to three (independent) translational Killing vectors. The curved FLRW metric \eqref{eq:curved_FLRW_small_k}, on the other hand, has six Killing vectors, half of which correspond to spatial rotations
\begin{equation}
    R_r^i = \epsilon_{rij}x^j\,,
    \label{eq:rotation_killing}
\end{equation}
and the remaining to spatial translations
\begin{equation}
    T_r^i = \delta^i_r - \frac{k}{2}x^i x^r + \frac{k}{4}\delta^i_r \mathbf{x}^2 + \mathcal{O}(\mathbf{x}^3)\,.
    \label{eq:translation_killing}
\end{equation}
These vectors have the following commutators:
\begin{align}
    [T_r, T_s] &= k\epsilon_{rst}R_t\,,\\
    [R_r, R_s] &= \epsilon_{rst}R_t\,,\\
    [R_r, T_s] &= \epsilon_{rst}T_t\,.
\end{align}
When $k=0$, the metric has the shape \eqref{eq:my_bianchi_I}. This has the three translational Killing vectors $\xi_r^i = T^r_i = \delta_r^i$ which commute among themselves. This is in fact the algebra of Bianchi I. 

The curved case is more elaborate. Indeed, solving the Killing equation to linear order in perturbations and expanding in powers of $\mathbf{x}$ can yield too many solutions. As a trivial example, consider the vectors $T_r \pm (\lambda k) R_r$ (where $\lambda$ is an arbitrary $\mathcal{O}(H^{-1})$ distance scale). They are Killing vectors of the perturbed metric at linear order in $k \sim \nabla^2\R$ and $\gamma_{ij}$, but clearly the constant-time metric is not that of a maximally symmetric space.

To make the comparison we rather ask whether there is some extension of the metric \eqref{eq:metric} to arbitrary order in $k$ (but still expanding in gradients) such that it is a Bianchi metric. Since anisotropy is small, we argue that this shows that \eqref{eq:metric} is indistinguishable from such a homogeneous but anisotropic universe.

For the closed $k > 0$ case, the solution to this question can be found in \cite{King:1991jd}. We reproduce their result here expanding to second order in $\mathbf{x}$. The Killing vectors of Bianchi IX are
\begin{equation}
    \xi_r^{\pm} = \frac{1}{2}(T_r \pm \sqrt{k}R_r)\,,
    \label{eq:xi_pm}
\end{equation}
where one chooses either the plus or minus sign for the three translational Killing vectors of this space. These vectors satisfy the commutation relations
\begin{equation}
    [\xi^\pm_r, \xi^\pm_s] = \epsilon_{rst}\xi^\pm_t\,,\quad [\xi^\pm_r, \xi^\mp_s] = 0\,.
\end{equation}
Note that $\xi^-_r$ are invariant under the transformations generated by $\xi^+_r$ and vice-versa, since $\mathcal{L}_{\xi^\pm_r}\xi^\mp_s = [\xi^\pm_r, \xi^\mp_s] = 0$. Now suppose that we choose $\xi^+_r$ to be the three translational Killing vectors. One can check from equations \eqref{eq:rotation_killing} and \eqref{eq:translation_killing} that
\begin{equation}
    \delta^{rs}\xi^-_r \xi^-_s = \left(1 + \frac{k}{2}\mathbf{x}^2\right) \delta^{ij}\,,
\end{equation}
so these vectors indeed form an invariant basis at this order: $e_r = \xi^-_r$. Similarly, if $\xi^-_r$ are chosen as the Killing vectors, $\xi^+_r$ form an invariant basis and satisfy a similar equation. Now we can explicitly build a metric which, to linear order in perturbations, has the same shape as \eqref{eq:metric} in the case of positive $k$, and which has three translational Killing vectors
\begin{equation}
g^{ij} = a^{-2}(t)\left(\delta^{rs} - \gamma^{rs}(t)\right)e_r^i e_s^j\,,
\label{eq:bianchiix_invbasis}
\end{equation}
where we have expressed the tensors in the invariant basis. Since $\gamma^{rs}$ are just numerical coefficients, this is explicitly invariant under the transformations generated by the Killing vectors. Note also that at leading order in $k$ and $\gamma_{ij}$ we have $\gamma^{rs}e^i_r e^j_s = \gamma^{ij}$, so to leading order, using equation \eqref{eq:bianchiix_invbasis}, this is
\begin{multline}
g^{ij} \approx a^{-2}(t)\left(\delta^{rs}e_r^i e_s^j - \gamma^{rs}(t)\delta_r^i\delta_s^j\right) \\= a^{-2}(t)\left[\left(1 + \frac{k}{2}\mathbf{x}^2\right)\delta^{ij} - \gamma^{ij}(t)\right]
\end{multline}
This is indeed equal to \eqref{eq:metric}. In this sense, Bianchi IX is a positively curved universe with a homogeneous tensor mode superimposed.

We can ask if something similar happens for the negatively curved case $k < 0$. Recall that, according to our discussion in Section~\ref{sec:einstein_comparison}, model V is compatible with FLRW perturbations if $a_i$ lies along the axis where $\gamma_{ij}$ is transverse; let us take this to be the $z$-axis. Then, the Killing vectors for Bianchi V can be written as
\begin{equation}
    \xi_A = T_A + \sqrt{|k|}\epsilon_{AB}R_B\,,\quad \xi_3 = T_3\,,
\end{equation}
where capital latin indices run in the range $[1, 2]$. These satisfy the commutation relations
\begin{equation}
    [\xi_1, \xi_2] = 0\,,\quad [\xi_A, \xi_3] = \sqrt{|k|}\xi_A\,.
\end{equation}
We can find a set of vectors invariant under the transformations defined by these Killing vectors by integrating the equation $\mathcal{L}_{\xi_r} e_s = [\xi_r, e_s] = 0$. We do this perturbatively in $\mathbf{x}$, and normalize the basis $e_r$ to be orthonormal to zeroth order in $\mathbf{x}$. We obtain after some work 
\begin{align}
    e^i_A &= \delta^i_A + \sqrt{|k|} \delta^i_3 x^A \nonumber\\
    &\phantom{=} - \frac{|k|}{2}x^Ax^B\delta_B^i +\frac{|k|}{2}x^Az\delta^i_3 - \frac{|k|}{4}\mathbf{x}^2\delta_A^i + \mathcal{O}(\mathbf{x}^3)\,,\\
    e^i_3 &= \delta^i_3 - \sqrt{|k|} \delta^i_A x^A \nonumber \\ 
    &\phantom{=} - \frac{|k|}{2} x^A z \delta^i_A + \frac{|k|}{2}z^2\delta_3^i - \frac{3}{4}|k|\mathbf{x}^3\delta^i_3 +\mathcal{O}(\mathbf{x}^3)\,.
\end{align}
With these expressions one can check that
\begin{align}
\delta^{rs}e^i_r e^j_s & = \delta^{ij} - \frac{|k|}{2}\mathbf{x}^2 \delta^i_A \delta^i_B \delta^{AB} - \frac{|k|}{2}\mathbf{x}^2 \delta^i_3 \delta^j_3 \\ 
& = \delta^{ij}\left(1 - \frac{|k|}{2}\mathbf{x}^2\right)\,.
\end{align}
Therefore, we can again write a metric which is explicitly invariant under the transformations generated by the Killing vectors and which, to leading order in perturbations, has the shape \eqref{eq:metric}, namely $g^{ij} = a^{-2}(t)(\delta^{rs} - \gamma^{rs})e_r^i e_s^j$. In this sense, Bianchi V is equivalent to an open FLRW universe with a homogeneous tensor mode superimposed. Finally, note that this result should not depend on the gradient expansion, since we can use an invariant basis without expanding in powers of $\mathbf{x}$ to construct this same metric and it should coincide with ours at the relevant order.

\section{Conclusions}
\label{sec:conclusions}

We studied the leading physical effect of superhorizon scalar and tensor perturbations of a FLRW universe with adiabatic evolution. We have shown that these are captured by three particular Bianchi solutions, namely, models I, V and IX.

Superhorizon perturbations can be expanded in gradients as in equation \eqref{eq:grad-expansion}. It was shown in \cite{Maldacena:2002vr, Creminelli:2004yq, Creminelli:2012ed, Hinterbichler:2012nm} that the leading and subleading orders in that expansion can be eliminated through a coordinate transformation. This assumes an adiabatic evolution such that to this order the time dependence of the perturbations can be ignored \cite{Bardeen:1980kt}. The leading physical effect is therefore given by the time-dependence of perturbations (which is suppressed by $\nabla^2/H^2$) and second order derivatives. Ignoring tensor fluctuations, it is well known that the leading physical effect of scalar perturbations is locally equivalent to adding a small curvature $k = -\frac{2}{3}\nabla^2\R$ (see e.g. \cite{Brandenberger:2004ix}). We rederived this fact starting from the comoving gauge and without discarding the time-dependence of the perturbations in section \ref{sec:laplacian_curvature}.

Less studied is the effect of a superhorizon tensor mode in the presence of spatial curvature. We show that accounting for second order derivatives and the time dependence of the scalar and tensor modes, one can choose coordinates to write the metric in a simple form given by \eqref{eq:metric}. This allowed us to show that the perturbed metric in the presence of tensor perturbations is locally equivalent to a Bianchi universe. The type of universe depends on the value of the curvature induced by the scalar perturbation $k$, such that models I ($k=0$), V ($k<0$) and IX ($k>0$) are recovered.

In section \ref{sec:einstein_comparison} we compared the Einstein equations and found that Bianchi solutions of types VII$_0$ and VII$_h$ cannot emerge as adiabatic perturbations of a FLRW universe with an untilted perfect fluid. In section \ref{sec:metric_comparison} we built the corresponding Bianchi metrics for the other cases. We used the result of \cite{King:1991jd}, who showed that Bianchi IX can be interpreted as a tensor mode superimposed on a positively curved FLRW universe, to interpret the $k > 0$ case. We then showed that Bianchi V can also be interpreted as a tensor mode superimposed on a negatively curved universe, to second order in the gradient expansion.

As a follow-up, it will be interesting to explore the observational consequences of these results, for example, by exploring whether they allow us to place an observational floor when constraining anisotropy at horizon scales, similar to what happens to spatial curvature~\cite{Leonard:2016evk}. From a theoretical perspective, it would also be interesting to explore the connection between Bianchi V and an open universe with a homogeneous tensor mode non-perturbatively, similarly to what is done in \cite{King:1991jd} for Bianchi IX.

\begin{acknowledgments}
J.N. is supported by Fondecyt Regular Grant No. 1211545. This grant also covered a collaboration visit by T.S.P. and gave additional support for S.R. T.S.P. is supported by FAPERJ (grant E26/204.633/2024), CNPq (grant 312869/2021-5) and Fundação Araucária (NAPI de Fenômenos Extremos do Universo, Grant No. 347/2024 PDI). S. R. was supported by Beca
Manutención de postgrado Magíster 2024 (PUCV).
\end{acknowledgments}

\appendix

\section{Solution to the constraint equations}
\label{sec:constraints}

In this Appendix, we derive the form of the lapse and the shift~\eqref{eq:lapse_shift_solution} from the constraint equations. The starting point is the metric in the ADM formalism \footnote{Note that we have factorized the scale factor. This means that quantities defined in terms of derivatives of $h_{ij}$ receive no background contribution.}
\begin{equation}
  ds^2 = -N^2dt^2 + a^2(t)h_{ij}(dx^i + N^idt)(dx^j + N^jdt)\,,
  \label{eq:metricADM}
\end{equation}
where $h_{ij} = (1 + 2\R)\delta_{ij} + \gamma_{ij}$. The momentum and Hamiltonian constraints are
\begin{gather}
  \!\!\phantom{.}^{(3)}\!R + K^2 - K_{ij}K^{ij} = 16\pi G \rho\,,\label{eq:hamilton_constraint}\\
  D_i(K^{ij} - a^{-2}h^{ij} K) = -8\pi G N T^{0j}\,,\label{eq:momentum_constraint}
\end{gather}
where the extrinsic curvature is $K_{ij} = \frac{1}{2a N}(\partial_t(a^2 h_{ij}) - D_i N_j - D_j N_i)$ and $\!\!\phantom{.}^{(3)}\!R$ is the three-dimensional curvature scalar computed from the metric $a^2 h_{ij}$.

The momentum constraint at linear order becomes
\begin{gather}
  4H\partial_i\delta N - 4\partial_i\dot{\R} + a^{-2}\partial_j\partial_i N_j - a^{-2}\nabla^2 N_i = 0\,,\label{eq:momentum_linear}
\end{gather}
where we have used the fact that in comoving gauge $T^{0i} = 0$. The transverse part of \eqref{eq:momentum_linear} gives $N_i^\perp = 0$, while the longitudinal part gives
\begin{equation}
\delta N = \frac{\dot\R}{H}\,.
\end{equation}
The Hamiltonian constraint is then, at linear order
\begin{equation}\label{eq:hamiltonian_linear}
  \frac{4}{a^2}\nabla^2\R  - 8Ha^{-4}\partial_i N_i = 16\pi G\delta\rho\,.
\end{equation}

It is easy to check from stress energy conservation that in comoving gauge that $\delta\rho$ is sourced by $\dot\R$ at linear order
\begin{multline}
    \dot\rho + \Gamma^i_{i 0} (\rho + p) = 0 \\ \implies \delta\dot\rho + 3H (1 + c_s^2) \delta\rho = - 3(1 + w) \bar{\rho} \dot\R\,.
\end{multline}
Therefore $N_i \sim \mathcal{O}(q/H)$.

\section{Einstein equations for a Laplacian of a scalar fluctuation}
\label{sec:einstein_laplacian}

We want to check that the Einstein equations for the metric \eqref{eq:curved_flrw_with_b} are those of a curved FLRW. Note that the time-dependence of $b(t) = a(t + \xi^0)(1 + \R(t))$ is fixed by the time-dependence of $a$, $R$, $\xi^0$ so we want to check that this time dependence solves the curved FLRW Friedmann equations.

Taking a time derivative of the effective scale factor $b(t)$ gives
\begin{multline}
\left( \frac{\dot{b}(t)}{b(t)} \right)^2 = \left(\frac{\partial_t a(t + \xi^0)}{a(t + \xi^0)} + \dot\R\right)^2 \\ = H^2(t+\xi^0)=\frac{8\pi G}{3}\bar{\rho}(t+\xi^0)\,,
\label{eq:bdot_over_b}
\end{multline}
where the last equality comes from the background Friedmann equation for the flat universe. The function $\bar{\rho}(t)$ is the background density of a flat universe. We want to connect this with the background density of the curved universe. For that, take the perturbed density in comoving gauge
\begin{equation}
    \rho_{co}(t) = \bar{\rho}(t) - \frac{1}{4\pi G a^2}\nabla^2\R\,.
\end{equation}
Note that $\rho_{co}$ depends only on time at this order in the gradient expansion. Now transform it to synchronous gauge
\begin{equation}
    \rho_{sync}(t) = \bar{\rho}(t + \xi^0) - \frac{1}{4\pi G a^2}\nabla^2\R\,.
\end{equation}
Therefore, we rewrite equation \eqref{eq:bdot_over_b} as
\be
\left( \frac{\dot{b}(t)}{b(t)} \right)^2 = \frac{8\pi G}{3} \rho_{sync} + \frac{2}{3b^2} \nabla^2 \R\,,
\ee
where we used the fact that $a = b$ at zeroth order in perturbations. This is indeed the Friedmann equation for a closed universe with $k = - \frac{2}{3}\nabla^2\R$.

\bibliographystyle{h-physrev4}
\bibliography{references}

\end{document}